# Time-resolved radiative recombination in black silicon

Seref Kalem[1,*]

[1] Department of Electrical and Electronics Engineering, Faculty of Engineering and Natural Sciences, Bahcesehir University, Besiktas, 34353 Istanbul, Turkey

## ABSTRACT

Black silicon (b-Si) has been receiving a great deal of interest for its potential to be used in applications ranging from sensors to solar cells and electrodes in batteries due to its promising optical, electronic and structural properties. Several approaches have been used to demonstrate the possibility of producing application quality b-Si, which also exhibits light emission properties. The photoluminescence is a useful technique to identify recombination pathways and thus, enable us to optimize device quality. In this work, we report the results of the radiative recombination dynamics in b-Si produced by a technique involving thermal oxidation, photoresist coating and chlorine plasma etching. An ultrafast blue luminescence component competing with non-radiative recombination at surface defects was identified as no-phonon radiative recombination. This component involves two decay processes with a peak energy at around 480 nm, which have the fast component of about 15 ps followed by a component of around 50 ps lifetime. The emission exhibits a slow process in red spectral region with time constant of 1500 ps. When the surface is smoothed, the lifetime of carriers increased up to 4500 ps and the emission peak blue shifted indicating downsizing in dimensions. The results are correlated with transmission electron microscopy, localized vibrational modes and spectroscopic ellipsometry and interpreted through the presence of quantum confinement at the tip regions of the wires, surface defects and oxide environment surrounding the nanoscale wires.

## 1 Introduction

Black silicon (hereafter b-Si) is an integral part of Si wafer and therefore, totally compatible with the microelectronics fabrication process. It can be selectively produced on a wafer and related metallization or surface passivation steps can be readily applied to it in order to realize any specific device for applications. b-Si, a special form of silicon having a surface structure consisting of quantum size wires, is formed as a result of surface etching using methods such as ultrafast laser pulsing of surfaces or reactive ion etching [1–7]. This form of Si has been the subject of significant attention due to its rich optical and electrical properties offering a great potential for promising applications for advanced scientific

Springer

research subjects [8–10]. For solar energy harvesting applications to generate photovoltaic effect, black silicon surface is a perfectly compatible material with silicon solar cells providing a natural advantage with its ease of fabrication advantage [11]. Solar cells fabricated using b-Si exhibited conversion efficiencies ranging from 10% up to 19% depending on the fabrication method and device structure [12–19]. Not only capturing the solar spectrum but also turning these wires into field effect devices can offer a great advantage for sensor applications [20]. Light trapping can enable the realization of the photothermic conversion devices [21]. Isolating particles of interest within the forest of quantum nanoscale wires or using surface functionalizing techniques enables us to enhance the sensitivity while providing an advantage of filtering the particles to be detected [8, 21]. Black silicon may find applications in imaging and micro-electro-mechanical systems (MEMS) as active and passive micro- and nanostructured semiconductor quantum structures [22–24]. The recent progress of black silicon ranging from fabrication to applications has been outlined in greater detail elsewhere [25].

The fabrication methods of b-Si include RIE or ICP-RIE, PIII, DRIE, pulsed laser surface treatment, Alkaline etching, stain etching, MACE, Electrochemical HF etching, FFC Cambridge process, etc. [14, 16 and references therein]. The reason of using our method of b-Si fabrication is simply based on the wider use of chlorine plasma for etching under class-1 clean room device production conditions. Particularly, the use in nanostructure formation and the production of feature sizes down to 10 nm and beyond technology nodes chlorine etching is required. The edge effects and black silicon formation during the deep trench etching of Si motivated us to produce b-Si on wafers [1]. Larger size of chlorine atoms as compared to Si could have generated b-Si like random etching structures on non-patterned blank PMMA/$SiO_2$/Si surface.

The surface formed on silicon wafers by above-mentioned techniques consists of tapered wires presenting quantum size structures down to a nanometer size tips [4]. It is interesting to see that the surface so obtained soaks all the visible light making the surface appear as a deep black color. Nevertheless, all those wires were found to be crystalline as we have already demonstrated through transmission electron microscope (TEM) analysis [4]. However, the surfaces of these wires are not atomically flat, representing some surface defects like structure, missing atoms and oxidation-induced capping regardless of wafer orientation. The presence of oxide cap layer or the oxidation of the surface of the nanoscale wires was also demonstrated by Fourier Transformed Infrared (FTIR) measurements of vibrational spectra. The presence of stretching modes of Si–O–Si vibrations at around 1090 $cm^{-1}$ provides a clear evidence of the oxide formation at the surfaces of the wires.

Time-resolved photoluminescence is a very important technique in assessing possibility of applications ranging from electronic to optical devices for any given semiconductor. It provides a very crucial information on carrier lifetimes through the excited states recombination dynamics. Analyzing the transient data, one can identify the radiative recombination processes in a bulk crystal or surface states. Hence, one can use this method in assessing defects, which would enable us to identify methods of optimizing quality of the b-Si for applications. There has not been a comprehensive and so detailed investigation of dynamical properties of the excited carriers in b-Si. Owing to its quantum size surface structure, we have already reported an efficient continuous wave (CW) photoluminescence (PL) activity in a broad spectral range going from visible to near infrared region [4]. CW PL exhibits wafer (p-type or n-type), crystal structure and temperature-dependent features both in the visible and in the near infrared region. In the literature, one can find a number of studies on time-resolved photoluminescence of silicon microstructures and nanostructures [26, 27]. Despite several studies on CW photoluminescence and time-resolved PL on micro- and nanostructures, there is no study on the radiative recombination dynamics of the quantum wires in black Si [28, 29]. This work aims at filling this gap of exciting scientific and technological interest in such a surface.

## 2 Experimental methods

### 2.1 Transformation of silicon surface to black silicon nanoscale wires

The b-Si quantum wires or whiskers were fabricated by an inductively coupled plasma-reactive ion etching (ICP-RIE) of thermally oxidized and photoresist (named as PMMA) coated 3-inch size *p*-type Si wafers having <100> and <111> crystal orientations as reported earlier [4]. Wafer resistivities used in the fabrication of b-Si in our work were around 10 $\Omega$-cm. For the RIE of these wafers, we used chlorine plasma, which has led to the formation of nanoscale wires on the surface of the Si wafer. Following this process, the wafer surface becomes black as appearance, since wires or whiskers absorb all the white light turning the surface of the wafer to a deep black color. Our method involving thermal oxidation, PMMA coating and chlorine plasma has shown that very sharp, tunable length and still crystalline nanoscale wires or pillars can be selectively fabricated on silicon wafers. Figure 1 shows the scanning electron microscopy (SEM) of the black silicon produced on **a** Si(100) and **b** Si(111) wafers. The energy-dispersive x-ray spectroscopy (EDX) indicates that the b-Si consists of silicon and oxygen atoms with the presence of about 4.30 atomic % of oxygen and 95.70 atomic % Silicon. For both of the samples, the presence of high density of individual pillars is typical, which is indicative of large surface area.

### 2.2 High-resolution transmission electron microscopy (HRTEM) images

TEM studies on these wafers have shown that wafer surface consists of Si quantum wires or whiskers type of extensions, which are aligned vertically to the wafer plane (Fig. 2a). As evidenced from the atom interference fringes observed in TEM as shown in Fig. 2b, these wires are crystalline and their dimensions can be down to few atoms thick tips. On their surface, some missing atom positions resulting in a certain irregular surface profile are visible. As shown in Fig. 2a, their length can be up to 200 nm long with some core–shell like structure having a crystalline Si core, which is encapsulated by a surface oxide [4]. Within the Si core, regular atomic planes with inter atomic distances of 0.34 nm can be measured and the surface irregularities due to missing atoms are also visible as shown in Fig. 2c. Particularly, the kinks with different sizes are clearly visible on the surface of the wires. We note that all the kinks have the same bending angles that is 105°. Nanoscale wire surfaces have some oxidation as we understand from the HRTEM image, but there is not a clear layered structure. Probably its average thickness is approximately within the low range of a nanometer. FTIR measurements in Fig. 3 indicate the existence of a native oxide on these wires as evidenced from a relatively strong Si–O–Si absorption band-related silicon–oxygen–silicon stretching vibrations at 1085 cm$^{-1}$.

The measurements made at different angle of incidences indicate also there is the presence of LO-TO coupling band [30] at 1254 cm$^{-1}$ as it can be seen in Fig. 3. The LO-TO coupling at Si–O–Si modes is an indication of the presence of structural disorder, which is also confirmed by the high-resolution transmission electron microscopy in Fig. 2. Although the b-Si pillars are single crystalline, their surfaces are not atomically smooth, representing for example some kinks along the surface.

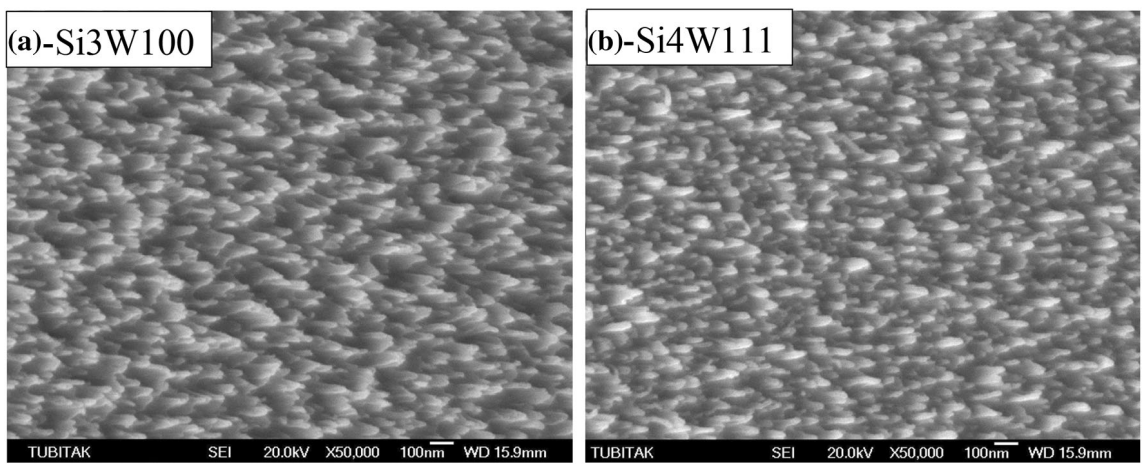

**Fig. 1** SEM of images of black silicon samples fabricated on **a** Si(100) and **b** Si(111) wafers. The scale is 100 nm as shown at the footer

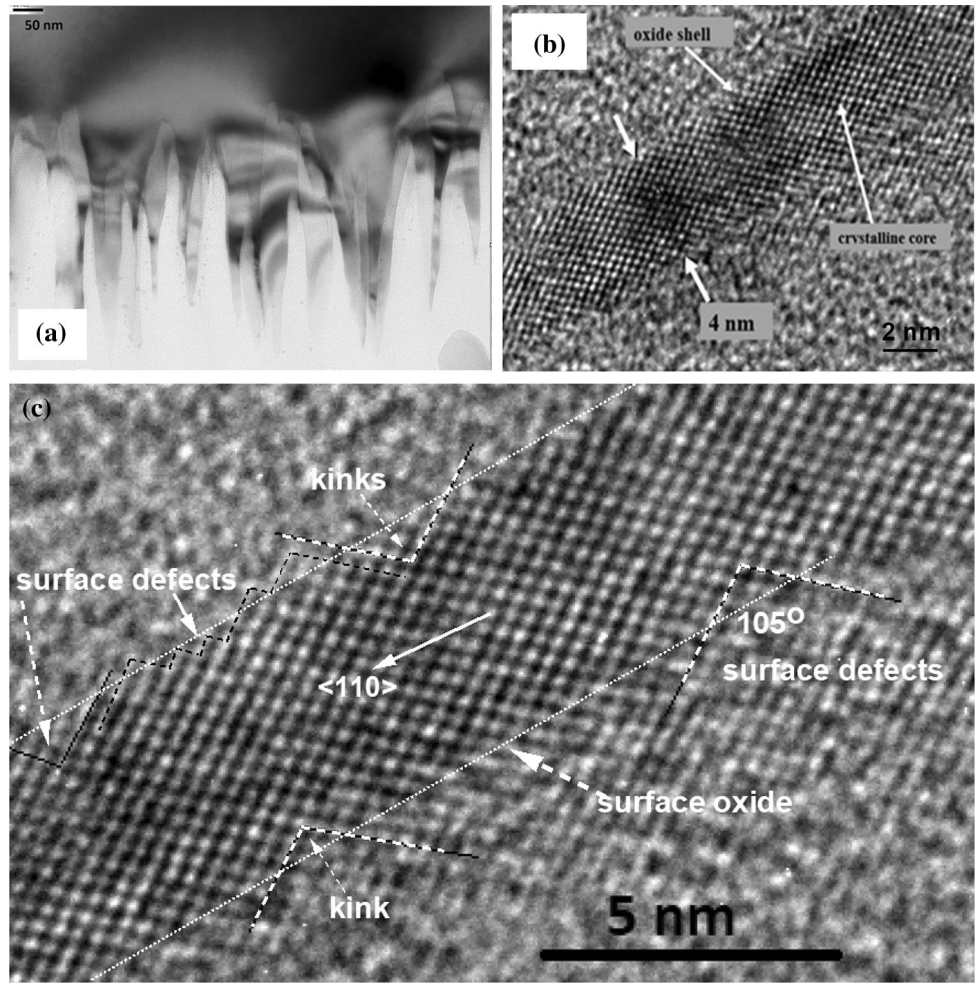

**Fig. 2** Transmission electron microscope image of black Si nanoscale wires. **a** TEM image of an individual wire, in which regular ordering of atoms and atom planes can be observed inside the crystalline structure. The surface of the quantum wire is indicative of defects consisting of some interrupted planes and missing atoms. The oxide shell indicates an oxide layer of about ∼ 1 nm on the wire surface. **b** TEM cross-sectional image at 50 nm magnification indicating an ensemble of quantum wires vertically aligned to the wafer plane (100). **c** Interface between the Si core and the oxide shell is separated with short length dots at both sides of the wire. Ordering of individual atoms and crystallographic orientation along the wire where the distances between atomic planes can readily be measured. The figure is typical of surface defects in the form of kinks as shown by angled lines (Color figure online)

## 3 Results and discussion

### 3.1 FTIR spectra and localized vibrational modes

As shown in Fig. 3, b-Si exhibits a strong absorption band at 1085 cm$^{-1}$, which originates from Si–O–Si symmetric stretching [31]. This band is indicative of the presence of an oxide layer on the surface of the b-Si quantum wires. From the close look at the surface details of a particular wire, we observe that the surface oxidation does not form a measurable layer in TEM. The remaining significant peaks in the spectra are due to Transverse Optical (TO) phonon modes of Si at 460 nm that is indicative of a disorder or amorphous like structure at the surface. The peak at 614 nm is the Si–Si phonon vibrations of the lattice. The insert shows the geometry of an individual silicon wire with the shaded tip region at the top where high energy photons are absorbed. The oxide layer surrounding of the wire is marked as red line capping the quantum wire structure. Note that there is no vibrational features relating to hydrogen–silicon bondings as evidenced from the absence of associated peak positions at around 2000–2000 cm$^{-1}$. Figure 3b displays the incidence angle-dependent transmission

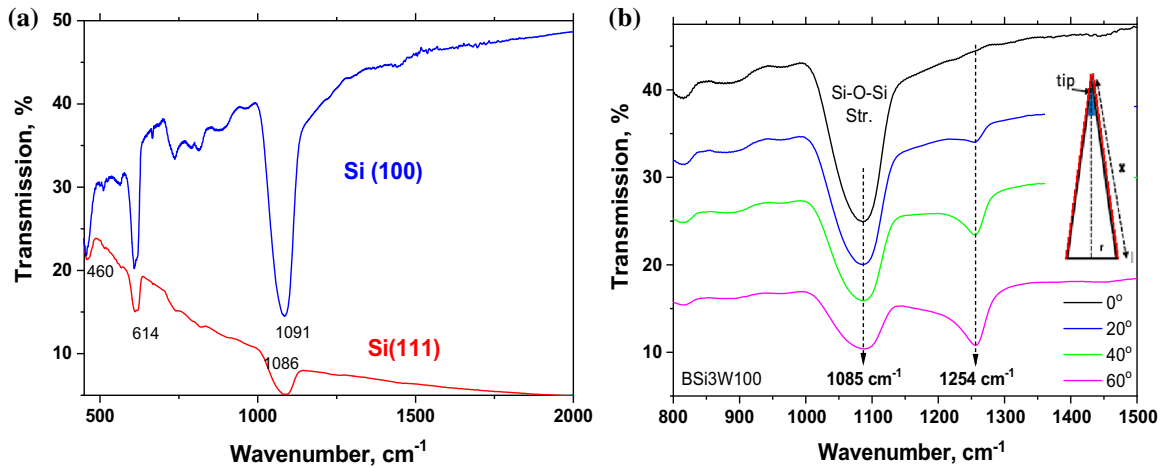

**Fig. 3 a** A typical FTIR spectrum of a black Si indicating the presence of oxidation as evidenced from the Si–O–Si stretching mode at 1085 cm$^{-1}$. **b** the incidence angle dependence of the Si–O–Si modes in Si(100) sample, revealing the appearance of some disorder coupled LO-TO phonon mode pairs at 1254 cm$^{-1}$ as reported earlier [30]. The insert shows the geometry of an individual silicon nanoscale wire with the shaded tip region at the top where high energy photons are strongly absorbed. The oxide layer surrounding of the wire is marked as red line capping the quantum structure (Color figure online) spectra of a b-Si. When the angle is increased, the strength of the band at 1085 cm$^{-1}$ decreases in favor of the new band at 1254 cm$^{-1}$. We attribute this type of behavior to disorder-induced splitting effects of LO-TO phonon modes as observed earlier [30].

### 3.2 Ultrafast radiative recombination dynamics and blue–green–red light emission

For the photoluminescence (PL) excitation dynamics, we used 100 femtoseconds Ti:Sapphire laser pulses at 400 nm as excitation wavelength and 80 MHz repetition rates. The power of the excitation laser pulses was ranged from 100 μW to 8mW. The signal detection system used a Streak camera for ultrafast PL decays and a Time-Correlated Single Photon Counting (TCSPC) measurement setups for the measurement of slow state decays.

Excited state dynamics of carriers ranging from blue to red spectral region were investigated by time-resolved photoluminescence (TRPL) at room temperature in b-Si consisting of high density quantum wires whose dimensions are less than the de Broglie wavelength as evidenced from TEM images [4]. An ultrafast decay component of about 10–15 picosecond (ps) and a second fast component of around 30–50 ps were deduced assuming a 3-components exponential decay process as measured by streak camera combined with TCSPC. The ultrafast PL decay results in a transfer of carriers to long-lived defect states as evidenced by a red emission at around 600 nm. Red shift at the initial stages of the blue luminescence decay confirms the presence of a likely charge transfer to long lived states. Time-correlated single photon counting measurements revealed a lifetime of about 1.5–2.5 ns (ns) for these states. We find that the same quantum structures emit also in the near infrared close to optical communication wavelengths. These results can be understood in terms of band structure modification at reduced sizes and defects induced at the surfaces involving dangling bonds and oxygen vacancies related radiative recombination centers in oxide on quantum wires. The nature of the emission can be described considering confinement and traps, and modeling can thus be provided for the excitation dynamics of charge carriers.

### 3.3 Streak camera image

Figure 4 is a typical wavelength-time 2D luminescence streak camera image of the emission from a b-Si consisting of a high density of nanoscale wires. Black-Si was produced on *p*-Si(111) wafer. Figure 4a and Fig. 4b are the images obtained on a sample before smoothing and after smoothing process, respectively. The streak camera image was taken using a laser excitation energy of 3.1 eV (400 nm). The apparent build up of the PL following the excitation is shown from 50 ps indicating a central intensity peak at around 490 nm. The temporal signal was attenuated very fast for times longer than 100 ps. Spectral intensity attenuates relatively fast at high energies and becomes weaker at longer wavelengths. For this sample, blue-green PL is typical without the presence of any other spectral component. The spectral evolution of the PL exhibits a clear intensity distribution over the time scale explored. The spectrum beyond the 550 nm is at the almost full attenuation regime. The 2D images presented in this article were obtained in relatively short times. The luminescence bands show a wide horizontal bands, while the laser gives a thin spot, indicating its short duration and narrow bandwidth. The data were used to provide reliable profiles for spectral and lifetime.

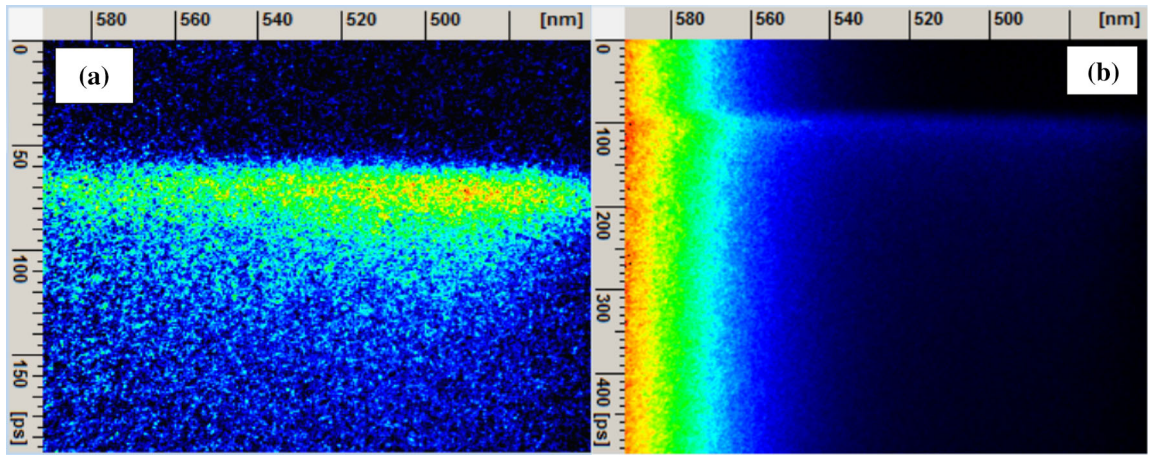

**Fig. 4** Streak camera image of PL emission. **a** PL intensity distribution as taken from streak camera image of the blue-yellow emission of a b-Si, which was photoexcited at 400 nm (3.1 eV) and recorded for 200 ps following excitation process. Red-yellow colored regions indicate the strongest location of time-resolved emission. **b** PL intensity distribution of b-Si wires, which were smoothed by exposing the surface to vapor of HF:HNO$_3$ acid mixture as explained elsewhere [32]. Note that in these images, the color scale ranges from dark to red corresponding to the highest emission intensity regions. The power levels off the excitation ranged from 100 μW to 8mW at an oblique incidence on the sample (Color figure online)

In order to determine the role of the oxide on the nanoscale wires, we have exposed the wires to the vapor of HF:HNO$_3$ acid mixture [32]. The results are summarized in Fig. 4b, which shows the streak camera image for the treated b-Si sample. As shown in the image, the radiative recombination line peaks at around 100 ps and decay very fast right away from the ultrafast emission line. The slower decay time in this surface smoothing process is probably indicative of the removal of defects on the surface of the wires as evidenced from the slower decays in ultrafast radiative recombination regime.

### 3.4 Recombination dynamics

In Fig. 5a and b, we observe the temporal and spectral evolution of the radiative recombination at 300 K on Si(111) sample, respectively. Figure 5a shows the amount of photon counting as a function of the decay time in picosecond at different wavelengths as deduced from Fig. 4. TRPL spectra were simulated by a curve fitting program resulting in three components decay times as typical dynamical behavior of b-Si samples. The following exponential expression was found to be well describing the carrier recombination dynamics:

$$I(t) = \sum_i A_i \exp\left[\left(-\frac{t}{\tau_i}\right)\right] \quad (1)$$

where $A_i (i = 1, 2, 3)$ represents the amplitude and $\tau_i$ is the decay time corresponding to each ultrafast component. The data are well fit by three exponential components: (around 13 ps), the decay time describing the radiative recombination at the Si core, recombination at surface defects or Si core-silicon oxide cap layer interface (44.5 ps) and a slow component (1450 ps) characterizing radiative decay time through oxygen-related defects within the surface oxide surrounding quantum wires, respectively. These lifetimes are actually much shorter than those observed in indirect bandgap bulk Si crystals. The reason for such short lifetimes is due to quantum

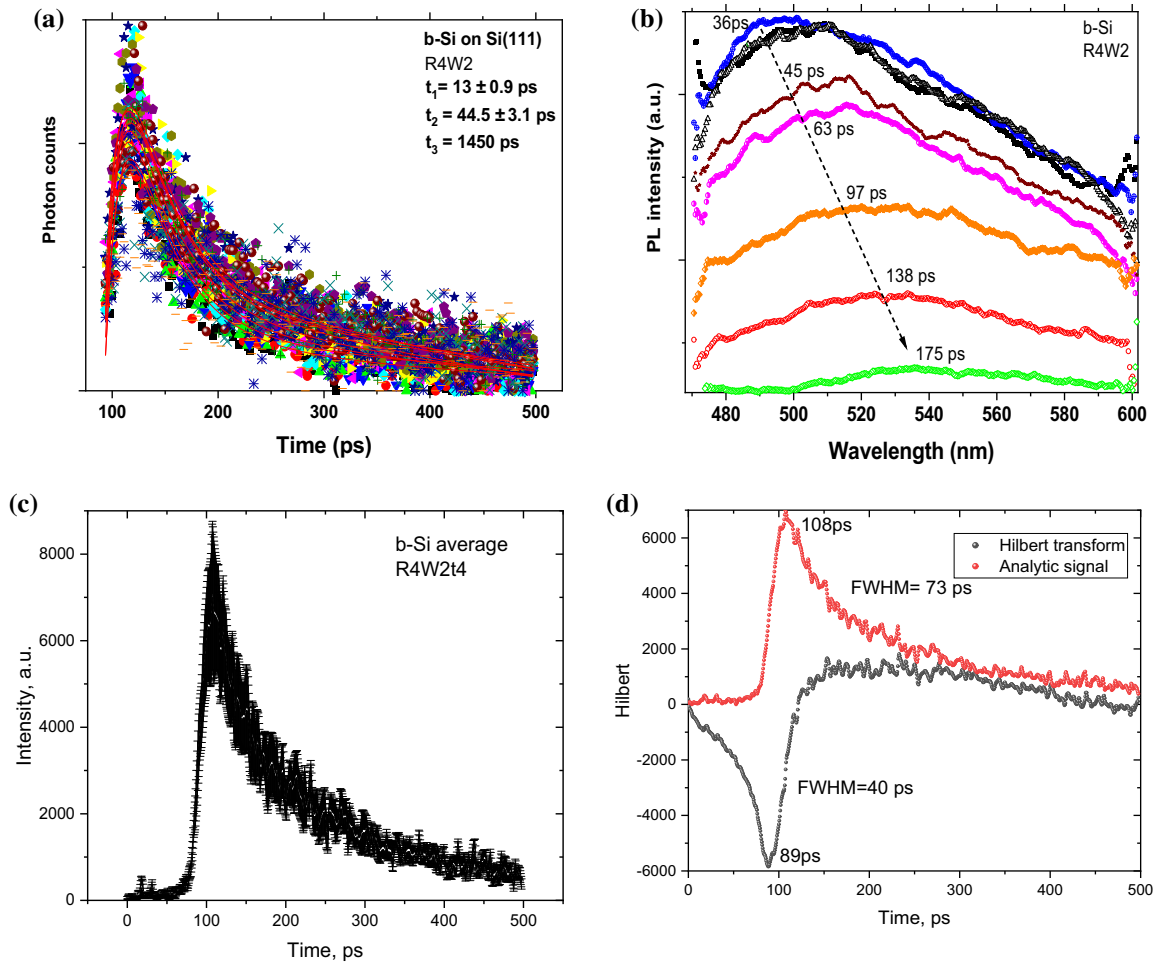

**Fig. 5** Room temperature excitation dynamics at 300 K. **a** Time-resolved PL spectra integrated from 450 to 600 nm b-Si. The solid red lines indicate a global fitting of the dynamics to a triple exponential decay functions. **b** Spectral distribution of PL between 470 and 600 nm taken at different radiative recombination decay times. As shown in Fig. 5b, the peak position of the time-dependent emission shifts toward the longer wavelengths, indicating the charge transfer to slower states (Color figure online)

confinement effects in the whiskers where the electron and the hole wavefunctions overlap within a very small volume. The second reason can be attributed to high defect density at the surfaces of the wires. The surface defects such as kinks as observed in Fig. 2 are indicative of the presence of a rough surface and the source of non-radiative points.

Figure 5c shows the intensity weighted average lifetime distribution determined by the statistical analysis. We use the Hilbert transform [33] to find out any hidden peaks in the spectra, and the result is shown in Fig. 5d. The Hilbert transform is related to the actual data by a 90-degree shift, and it is to extract analytic representations of signals from real valued measurements. It is given by the following expression

$$y(x) = f(x) + iH[f(x)] \qquad (2)$$

where $f(x)$ and $iH$ are the real and imaginary parts of the analytic signal and $x$ is the phase of the analytic signal. Thus, it is possible to separate the complex signal into the real and imaginary components. The transform indicates there is well defined a sharp peak at around 89 ps, which is in agreement with the average lifetime of 90 ps measured on a number of samples.

However, we cannot rule out the possibility of attributing the decay component of 44.5 ps to Auger recombination effect [34]. This effect has already been shown to be effective at low dimensional quantum structures at high temperatures under high-excitation conditions. Therefore, it is expected that free carrier generated under high excitation conditions can lead to secondary relaxation, decay component that is measured in our samples. However, the nanoscale wires are burned in to flames under high density of excitation; therefore, minimal excitation powers have been used to avoid the fire.

In order to verify the effect of any quantum confinement, we assume a nanowire or quantum dot-like confinement structures, which may be responsible for the ultrafast component of the light emission at around 480 nm. These dots can be expected to be located at the tips of the wires as evidenced from the TEM results. Following the analysis made previously [35–38], a quantum dot structure can confine electrons in three directions within a volume of $d_1$, $d_2$, $d_3$. The confinement energy can then be expressed in the following form:

$$E = E_c + E_{q1} + E_{q2} + E_{q3} + \frac{\hbar^2 k^2}{2m_c}$$

where $E_{qi} = \frac{\hbar^2 \pi^2}{2m_c} \left(\frac{q_i^2}{d_i^2}\right)$ with $q_i$=1, 2, 3, quantum number and $d_i = d_1, d_2, d_3$ are the unit length of the box in three dimensions and k is the vector component.

Assuming $d_1 = d_2 = d_3$ and $k = 0$, $E = E_c + 3 E_q$ where $E_q = E_{q1} = E_{q2} = E_{q3}$ and $E_c = 1.1$ eV as the bandgap of the bulk Si. $E_q = 2.89 \times 10^{-18}$ eV.m$^2$ (1/d)$^2$ where $d$ is in nm, $E_q$ is the confinement energy. From these estimations, we find that the ultrafast component of the emission is probably originating from the Si volume size of about $d = 2.5$ nm, that is 15.6 nm$^3$. Size-dependent change of the confinement energy is shown in Fig. 6b as an insert.

### 3.5 Smoothing of nanoscale wire surfaces

It has already been shown that the exposure of silicon rods to the vapor of an acid mixture containing HF and HNO$_3$ has the effect of smoothing the surface of the rods [32]. Similar effects are expected for the b-Si quantum wires. Therefore, the PL decay measurements have been performed also on b-Si wire samples, which were subjected to the surface smoothing procedure as described earlier [32]. This procedure removes some of the oxide from the shell region on the wire surfaces. The results obtained from these investigations on such samples are displayed in Fig. 6 for a b-Si quantum wire formed on p-Si(100). The findings are indicative of an increase in carrier lifetimes in b-Si wires, which have been subjected to surface smoothing process. This process removes the surface oxide but at the same time initiates a new oxide growth on the surface. This process can be interpreted as a surface smoothing removing of defects such as kinks while eliminating traps at the interface between the shell and the oxide layer encapsulating silicon wires. As shown in Fig. 6a, the stretched PL decay curves of etched nanoscale wires resulted in rather longer time-constants that is 12.7 ps, 30.9 ps and a slower component of 4500 ps as compared to as-fabricated b-Si quantum wires. This effect can be attributed to the fact that the surface smoothing passivates some of the traps at the surfaces and the enhancement of oxidation increases the lifetime of the carriers.

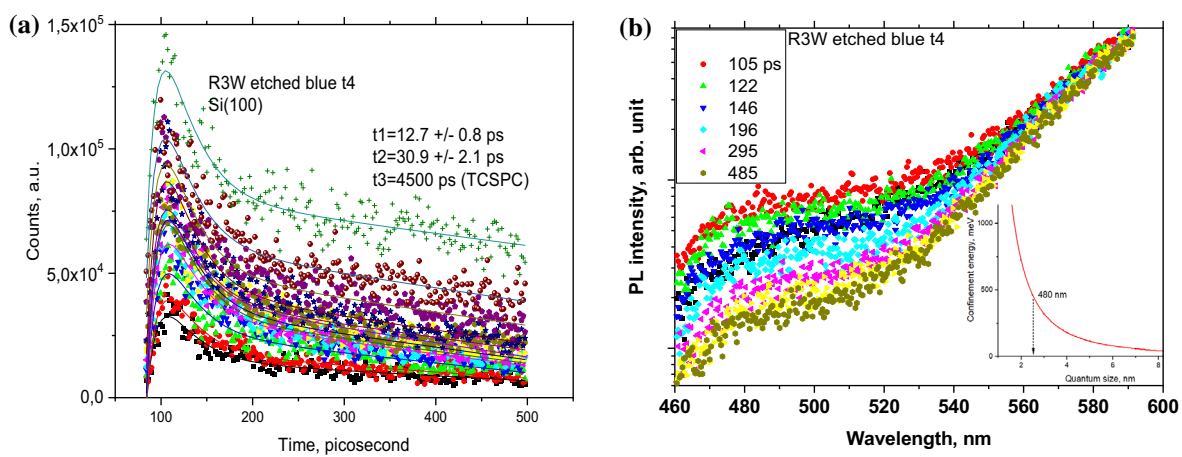

**Fig. 6** Room temperature excitation dynamics at 300 K for nanoscale wires treated by HF:HNO$_3$ vapors. **a** Time-resolved PL stretching spectra integrated from 450 to 600 nm of b-Si. The solid red lines indicate a global fitting of the dynamics to a triple exponential decay functions. **b** Spectral distribution of PL spectra between 470 and 600 nm as taken at different times (Color figure online)

Surface smoothing indicates that the surface recombination is more important than inside core and interfaces. Thus, the lifetime rise can be attributed to the passivation of surface defects by smoothing process.

Also, this effect results in the blue shift of the PL spectrum by about 5%, which is indicative of downsizing in pillar dimensions.

### 3.6 Time-correlated single photon counting

Figure 4 shows the time-resolved PL dynamics spectrally integrated from 450 to 600 nm for a black-Si produced on Si(111) as a function of decay time. The solid lines indicate a global fitting of the PL dynamics to a triple exponential decay functions. As indicated in Fig. 4, the blue emission at around 480 nm corresponds to the fast decay component of about 10 ps timeframe. This component could be attributed to direct transitions within the quantum wires due to the size reduced enhancement of the overlapping of the electron and hole wave functions leading to strongest oscillator strength. The fitting leads also to the presence of a component in the order of few nanoseconds corresponding to slow states induced probably by surface defects such as Si vacancies in the oxide layer surrounding quantum wires. An intermediate decay time of about 50 ps could be due to re-excitation of thermalized carriers or recombination in non-radiative traps at the interface between the silicon core and surface oxide layer. TCSP measurements revealed decay times ranging from around 1.5 ns up to 2.5 ns time range, which are attributable to vacancies in oxide layer.

Non-radiative recombination at traps was shown to be playing a significant role in photoluminescence decay in semiconductors [4]. Therefore, the intermediate decay component of around 50 ps could be attributed to the presence of traps at the surfaces of nanoscale wires. Radiative or non-radiative recombination at these traps is expected to play a significant role in determining the magnitudes of the decay time components in b-Si. While the non-radiative recombination reduces the emission intensity, the radiative one contributes to the spectral region between the blue and the red region.

From these results, we observe that the intensity of light emission at around 480 nm resulting from the ultrafast decay is rather weak. The reason of this can be the following: (1) quantum size concerns only the tip regions of relatively small volume as compared to the whole wire volume, (2) electron–hole pair generated at the tips of the quantum wires diffuses fast away from the tip regions to trap states or larger dimensional parts of the silicon core thus reducing the rate of the quantum confined radiative recombination of carriers. These processes are likely at the origin of the weak emission at around 480 nm. The most of the excitation light is absorbed in the remaining parts of the wires thus contributing light emission originating from the slow states at around 600 nm. A typical shape of a wire can be assumed to be a cone with a height of about 250 nm and with a cone diameter of around 100 nm as shown at the insert in Fig. 1d. The surface area of such a wire can be estimated from A = $\pi r x$, where $r$ and $x$ are the radius and the length of the lateral surface, respectively. Using these parameters, the surface area of the wire can be estimated to be $A \approx 4 \times 10^4$ nm$^2$ as compared to a tip region of 5 nm in size having a surface area of around 100 nm, leading to a ratio of surface area of about 400. This comparison indicates that actually a very small part of the light excitation occurs within the quantum sized region of the tip of the wire. This explains why the photoluminescence intensity is weak and the decay times are fast at high energies corresponding to tip regions where quantum size effects are expected. Beside the small volume of the tip regions, diffusion of generated electron–hole pairs from these regions toward the bulk part of the wires plays also a significant role in the strength of the high energy component of the time-resolved PL.

As shown in Fig. 7a, the PL peak position shift rate $d\lambda/dt$ is much slower in etched b-Si than as-fabricated one, which is indicative of a very fast charge transfer in b-Si. These findings suggest that the acid vapor treatment removes fast states, which are likely more effective at the tip regions of the wires, where quantum confinement effects are dominant. Secondary effect of the vapor treatment is to enhance the oxidation around the core silicon wire. The enhancement of the PL intensity of the long lived states in etched samples supports this type of evolution in recombination dynamics. Due to variable peak positions ranging from 470 to 540 nm, we attribute this band to hot carrier non-phonon emission originating from the core silicon that is the tip region of the wires. Thus, the short-lived component of about 5–15 ps could well be associated with the lifetime of these carriers, which are fast recombined at the interface traps between the shell and the oxide surrounding. These fast recombination processes can be modeled assuming a decay at the silicon core, oxide and the interface between the oxide and the core. This is schematically described in Fig. 7b as energy band diagram indicating the involvement of three states taking part in the recombination kinetics. The left hand side is the silicon core where the blue-green emission at around 2.5 eV occurs due to the confinement effect; the middle part indicates traps involved recombination due to localized states between the core and the surface oxide; the right hand side is the red recombination originating from the defects in oxide around the nanoscale wires.

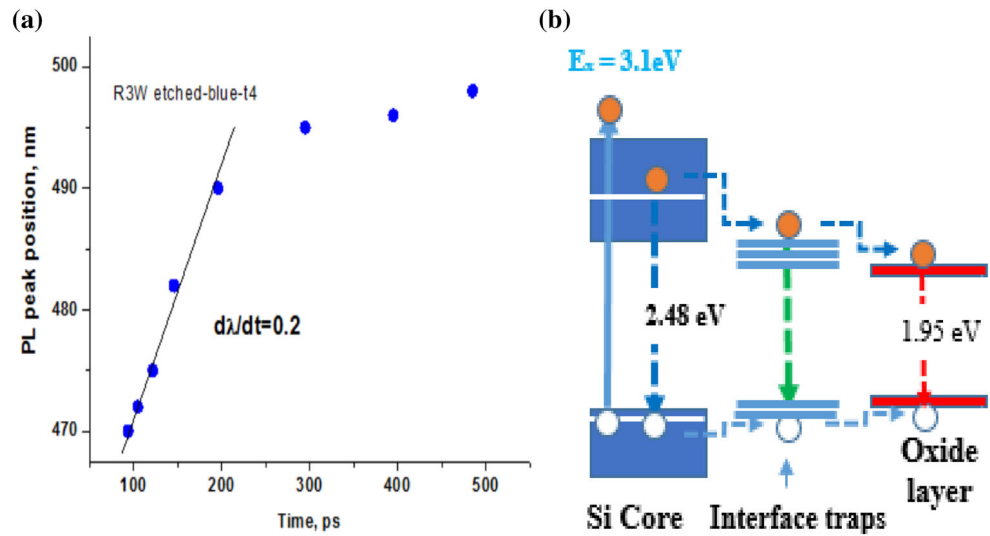

**Fig. 7** First derivative of the peak position versus decay-time. **a** PL peak position as a function of decay time exhibits two slopes for the change. The change rate $d\lambda/dt = 0.2$ over fast states is much higher than that for the slower states that is $d\lambda/dt = 0.019$, which is the typical characteristic of the slower states. **b** Energy band diagram indicating the involvement of three states taking part in the recombination kinetics. The left hand side is the silicon core where the blue-green emission at around 2.5 eV occurs due to the confinement effect; the middle part indicates traps involved recombination due to localized states between the core and the surface oxide; the right hand side is the red recombination originating from the oxide around the wires (Color figure online)

These results are well correlated with the continuous wave (CW) photoluminescence results in the visible and near infrared region as shown in Fig. 8. The tail emission at around 500 nm in Fig. 8a corresponds to the spectral distribution that we observe in the time-resolved studies. The strong red CW band proves the involvement of these states in the lifetime dynamics. The charge transfer from the fast states to slow states that we find in the time decay measurements favors the emission of this band. The assignment of the visible emission spectrum in these nanostructured silicon was attributed to quantum confinement effects in nanosize crystals of 2 to 3 nm in S-band and F-band corresponding to red and blue-green region, respectively [39]. Whereas, the visible emission is not the only emission type in the black silicon. If we use Germanium Photo Diode detector and the excitation source of 514 nm of Argon ion laser, an emission band at 1138 nm and a stronger band at 1525 nm are of typical peaks (Fig. 8b). The first band is characteristic band edge emission of the silicon, while the latter is related to defect emission, namely D1 type dislocations introducing deep levels into the material [40]. These emissions were observed at room temperature, and their intensities and distributions are dependent on the black silicon formation conditions. These dislocations play the role of recombination centers reducing the lifetime of the excited carriers. In addition to these emissions, it has been shown that the Terahertz emission was possible from the black b-Si fabricated by RIE [41].

## 4 Conclusions

Radiative recombination dynamics of the black silicon consisting nanoscale wires or whiskers have been investigated systematically on a number of samples in order to identify the origin of the radiative

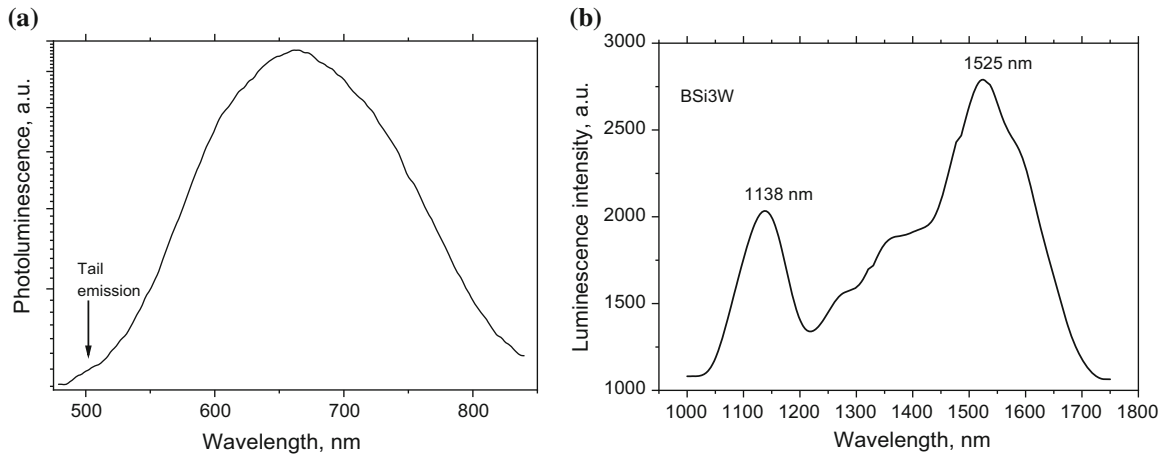

**Fig. 8** CW photoluminescence in black silicon in visible with a broad red emission band at around 650 nm (**a**) and near infrared region photoluminescence (**b**) at room temperature (Color figure online)

recombination process. An ultrafast decay component of excited carriers at around 480–500 nm with about 10–15 ps lifetime, contributing to a blue-green emission close to femtosecond regime is due to non-phonon electron–hole recombination within the nanoscale wires. This radiative decay component is followed by a slow recombination component in red spectral region, indicating a very fast charge transfer of carriers in black silicon nanoscale wires. The slow recombination lifetime can range from 1.5 ns to 2.5 ns as evidenced from the time correlated single photon measurements. Recombination component at intermediate time scales of around 30–50 ps can be attributed to trap states or defects where non-radiative recombination is a dominant effect at the interface between the core and the surface oxide of the nanoscale wires. We have shown when the surfaces of the b-Si nanoscale wires are smoothed by acid vapor treatment, the lifetimes increase, which is indicative of a decrease in surface defect states. The outcomes of these studies enable us to understand the underlying nature of radiative recombination kinetic in black silicon samples consisting of well aligned nanoscale wires. We have shown that the optical properties of b-Si can be controlled and tuned to meet the application requirements such as chemical and biological sensors, heat dissipation components, switching devices, optical filters, imaging, energy harvesting, and optical information processing platform integrating near infrared optoelectronic components.

## Acknowledgements

I thank Prof. W. Sunström for the access to Lund Laser Center and Dr. Hannas for the assistance in lifetime measurements. We thank Dr. P. Werner for HRTEM measurements and fruitful discussions.

## Author contributions

SK designed the experiments, prepared, characterized samples and analyzed the results and wrote the manuscript.

## Funding

I received infrastructure use funding from LASERLAB-EUROPE program under LLC001765.

## Data availability

The datasets generated during and/or analyzed during the current study are available from the corresponding author on reasonable request.

## Declarations

**Conflict of interest** The author has no competing interests to declare that are relevant to the content of this article.